# Livia: An Emotion-Aware AR Companion Powered by Modular AI Agents and Progressive Memory Compression

Rui Xi[1][0009-0004-9504-3085] * and Xianghan Wang[2][0009-0007-3437-3373] *

[1] University of California, Berkeley, Berkeley CA 94702, USA
[2] New York University, Brooklyn, NY 11201, USA
xr@berkeley.edu, xianghansharon@gmail.com
* These authors contributed equally to this work.

**Abstract.** Loneliness and social isolation pose significant emotional and health challenges, prompting the development of technology-based solutions for companionship and emotional support. This paper introduces Livia, an emotion-aware augmented reality (AR) companion app designed to provide personalized emotional support by combining modular artificial intelligence (AI) agents, multimodal affective computing, progressive memory compression, and AR-driven embodied interaction. Livia employs a modular AI architecture with specialized agents responsible for emotion analysis, dialogue generation, memory management, and behavioral orchestration, ensuring robust and adaptive interactions. Two novel algorithms—Temporal Binary Compression (TBC) and Dynamic Importance Memory Filter (DIMF)—effectively manage and prioritize long-term memory, significantly reducing storage requirements while retaining critical context. Our multimodal emotion detection approach achieves high accuracy, enhancing proactive and empathetic engagement. User evaluations demonstrated increased emotional bonds, improved satisfaction, and statistically significant reductions in loneliness. Users particularly valued Livia's adaptive personality evolution and realistic AR embodiment. Future research directions include expanding gesture and tactile interactions, supporting multi-user experiences, and exploring customized hardware implementations.

## 1 Introduction

Loneliness has reached epidemic levels globally, with nearly half of all adults in the United States reporting significant loneliness in recent years [1]. Chronic loneliness and social isolation pose severe health risks, including heightened depression and increased mortality rates comparable to smoking 15 cigarettes per day [1]. In response, researchers and practitioners have increasingly turned to technological interventions, notably

AI-driven companion apps or virtual agents, as accessible tools to provide emotional support, conversation, and companionship.

AI companions such as Replika have grown popular among individuals coping with loneliness, frequently serving as friends or informal mental health supporters [2]. A recent study involving 1,006 young adults using Replika showed that while these users experienced significantly higher levels of loneliness compared to average peers, they still perceived substantial social support from their AI companion [3]. Remarkably, 3% of users reported that interaction with the AI companion prevented suicidal ideation [3]. These findings underline the potential for thoughtfully designed AI companions to meaningfully impact emotional and mental well-being.

However, despite these promising outcomes, current virtual companions exhibit considerable limitations. Most commercial chatbots remain emotionally superficial, relying heavily on scripted responses or simple pattern matching, thus lacking genuine empathy or deep personalization [2,3]. These companions typically possess limited or no capability for long-term memory, resulting in repetitive and contextually insensitive interactions. Additionally, emotional state detection is often restricted to basic textual sentiment analysis, neglecting important multimodal cues such as voice tone. Further, most AI companions present a fixed or slowly evolving personality, limiting their ability to resonate deeply and adaptively with individual users. Over time, such superficial interactions may fail to establish sustained, meaningful relationships, ultimately undermining the companions' effectiveness.

To address these critical challenges, we introduce Livia, an emotion-aware augmented reality (AR) companion integrating advanced affective computing, modular multi-agent architectures, efficient long-term memory management, and interactive AR experiences. Our system aims to deliver personalized emotional support—perceiving emotions, recalling personal context, and dynamically adapting interactions with empathy.

Our key contributions include:
- Novel Modular Architecture: A robust, extensible multi-agent AI framework designed specifically for emotional and conversational interactions.
- Innovative Memory Management: Progressive compression methods that retain critical emotional and contextual information within practical constraints.
- Real-time Multimodal Emotion Recognition: Integration of multiple data streams enabling responsive and empathetic user engagements.
- Dynamic and Adaptive Personality: An evolving personality model that customizes interactions based on continuous user feedback.
- Immersive AR Integration: Enhancing user presence and emotional connection through expressive, embodied AR interactions.

## 2 Related Work

### 2.1 Emotional Virtual Agents and Memory Modeling

Our research extends prior work on emotional virtual agents, focusing on recognizing, interpreting, and responding to human emotions effectively. Empathetic dialogue systems trained on datasets like EmpatheticDialogues enable emotionally resonant chatbot responses [4], while embodied conversational agents and social robots demonstrate significant interaction improvements through facial expression recognition and affective reasoning [2]. Existing systems, however, struggle to maintain realistic long-term emotional connections due to limited memory capabilities, typically relying on simplistic dialogue summarization or explicit fact storage. Advanced models, such as the Generative Agents framework, introduced comprehensive memory management techniques, yet these methods come with considerable computational overhead, making them challenging to implement practically in resource-constrained environments [5].

### 2.2 Multimodal Affective Computing

Multimodal affective computing has emerged to overcome the limitations of single-modal emotion detection by integrating multiple emotional signals—such as facial expressions, voice tone, and textual sentiment analysis. Research indicates that combining modalities substantially enhances emotion recognition accuracy [6]. Livia leverages multimodal affective computing by integrating textual sentiment analysis and voice tone detection, providing a more robust and reliable emotional inference capability compared to single-modality approaches.

### 2.3 AI Companions in AR/VR

Augmented Reality (AR) and Virtual Reality (VR) technologies significantly enhance AI companionship experiences by embedding virtual companions directly into user environments, fostering deeper engagement and presence. Previous AR prototypes involving virtual pets or coaching agents demonstrated increased user satisfaction and perceived naturalness compared to traditional screen-based chatbots [7]. Essential to successful AR companionship is situational awareness, including realistic gaze alignment and real-time responsiveness to user behaviors. Livia builds upon this foundation, employing Unity3D and ARKit to integrate expressive avatars into users' real-world settings, thus elevating immersion, emotional bonding, and interaction authenticity through synchronized expressions and gestures.

### 2.4 Agent-Based Architectures and Behavioral Orchestration

Modern AI companions increasingly adopt multi-agent system architectures, decomposing cognitive tasks into specialized, independently operable components for improved modularity and performance [4]. Frameworks such as HuggingGPT and Auto-Gen exemplify this approach by utilizing orchestrator agents that delegate complex tasks across various specialized agents effectively [8]. Livia implements a similar modular multi-agent design, featuring dedicated agents responsible for emotional analysis, conversational interaction, memory compression, and system orchestration. This architecture promotes extensibility and seamless component integration, ensuring coherent, contextually rich user interactions while facilitating independent enhancements of individual agent capabilities.

## 3 System Architecture Overview

Livia employs a modular AI architecture that integrates seamlessly with an augmented reality (AR) frontend to create immersive and emotionally engaging user experiences. Fig. 1 illustrates the overall system structure, highlighting the integration between the AI agent stack and the AR rendering pipeline.

The backend features a multi-agent system comprising four specialized AI modules:

- **Emotion Analyzer Agent:** Responsible for assessing the user's emotional state by analyzing multimodal data, including textual sentiment and vocal intonation. This agent classifies emotions such as happiness, stress, sadness, or neutrality, providing critical emotional context for downstream interactions.
- **Frontend Voice Interaction Agent:** Utilizing a GPT-4-based conversational model, this agent generates personalized, emotionally appropriate dialogues. It synthesizes user history, emotional context, and memory content to dynamically craft responses, supporting both text-to-speech synthesis and pre-recorded audio interactions.
- **Memory Compression Agent:** Manages and compresses long-term memory through innovative algorithms—Temporal Binary Compression (TBC) and Dynamic Importance Memory Filter (DIMF). These algorithms enable efficient retention and retrieval of emotionally salient and contextually important interactions, ensuring continuity and depth in user relationships.
- **Behavior Orchestration Agent:** Functions as the central coordinator of the system, orchestrating the interaction among all specialized agents. It determines workflow execution by managing inputs, updating shared emotional and memory states, and ensuring coherent, contextually responsive interactions.

The frontend application, developed with Unity3D and Apple ARKit, renders an expressive 3D avatar into the user's physical environment. This AR interface enhances the sense of presence and social interaction through synchronized voice, gaze, and expressive animations driven by the backend's emotional state assessments and conversational outputs. The integration of AR with the modular AI backend provides a compelling, immersive user experience that effectively addresses emotional and companionship needs.

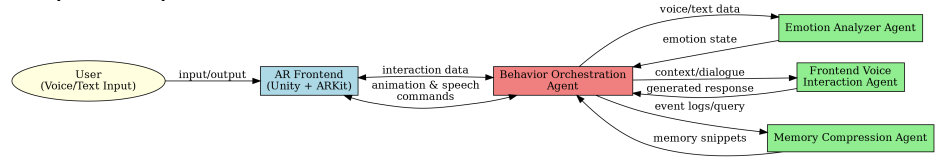

**Fig. 1.** Livia System Architecture

## 4 Modular AI Agent System

Livia's intelligence is structured as a modular multi-agent system (Fig.1), where each agent independently handles a distinct cognitive task, enabling robustness, scalability, and maintainability. Below we detail each agent's responsibilities and implementation:

### 4.1 Emotion Analyzer Agent

The Emotion Analyzer Agent continuously monitors and interprets the user's emotional state using multimodal data, including textual sentiment and voice intonation. For text analysis, we employ a RoBERTa-based classifier fine-tuned on emotion-labeled dialogue datasets, recognizing emotional states such as happy, sad, angry, anxious, and neutral. Voice emotion recognition utilizes a CNN-LSTM-based model, mapping audio inputs to valence and arousal dimensions and calibrating against the user's neutral baseline. Emotional states and confidence scores derived from these analyses are recorded in shared memory, influencing subsequent interactions and communication strategies.

### 4.2 Frontend Voice Interaction Agent

This agent generates natural, context-aware conversational responses utilizing a GPT-4-based language model accessed via the OpenAI API. It synthesizes multiple contextual inputs, including recent conversation turns, retrieved memory snippets, current emotional states, and user-selected personality archetypes (Fire, Water, Earth). Responses are dynamically adapted based on prompt prefix tuning to reflect appropriate emotional tones and interaction styles. The output is structured as plain text with metadata, guiding the AR frontend on vocal expression and animations to enhance the conversational experience.

### 4.3 Memory Compression Agent

The Memory Compression Agent maintains and compresses long-term interaction data using a lightweight SQLite database. Memory entries include timestamps, utterance types, emotional contexts, and importance ratings. Two novel algorithms manage data effectively:

- **Temporal Binary Compression (TBC):** Periodically compresses older interaction data hierarchically, merging detailed entries into concise summaries to efficiently reduce storage requirements.

- **Dynamic Importance Memory Filter (DIMF):** Activates when memory usage approaches predefined thresholds, prioritizing retention based on emotional intensity and contextual relevance, removing or summarizing less critical information. Relevant memories are retrieved through keyword and semantic similarity searches, providing historical context for ongoing interactions.

### 4.4 Behavior Orchestration Agent

This meta-agent serves as the system's central decision-maker, coordinating the flow of information and agent interactions. It uses a Python-based hybrid model combining explicit rule sets and reinforcement learning from user feedback to proactively manage user engagement strategies. The Orchestration Agent triggers interactions based on real-time emotional states, historical context, and overall engagement policy. It evaluates responses from other agents, ensuring coherence, appropriateness, and timely interactions. Additionally, it implements tone and politeness classifiers to maintain conversation quality and user comfort.

## 5 Progressive Memory Compression

A significant innovation in Livia is its advanced handling of long-term memory using two novel algorithms: Temporal Binary Compression (TBC) and Dynamic Importance Memory Filter (DIMF). These algorithms collectively ensure the scalability, responsiveness, and personalization of memory management, reflecting human-like memory retention and forgetting.

### 5.1 Temporal Binary Compression (TBC)

TBC systematically compresses memory based on temporal intervals, emulating human-like memory decay. Recent interactions are retained in fine detail, while older memories are progressively condensed into higher-level semantic summaries. The TBC algorithm operates hierarchically, merging memory entries pairwise within defined exponential time intervals (e.g., daily, weekly, monthly).

Specifically, TBC involves the following steps:
- **Epoch Definition**: Memory intervals are structured into levels with exponentially increasing durations—recent intervals remain detailed, while older intervals are coarser.
- **Pairwise Summarization**: Consecutive memory entries within each interval are merged into semantic summaries, effectively condensing detailed logs into concise descriptions.
- **Hierarchical Merging**: Summaries themselves are further merged at higher levels, continuously reducing memory granularity while preserving essential semantic content.

This hierarchical compression ensures efficient memory management, allowing the system to retain important contextual information without overwhelming memory storage (Fig. 2).

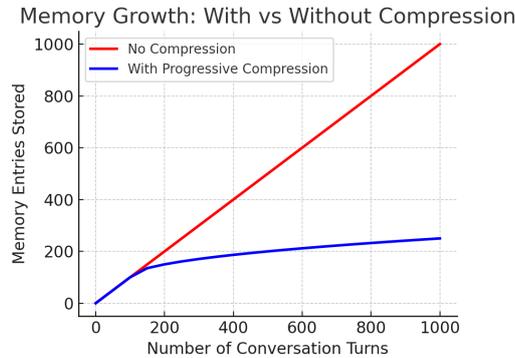

**Fig. 2.** Memory Growth with and without Compression

### 5.2 Dynamic Importance Memory Filter (DIMF)

Complementing TBC, DIMF dynamically manages memory storage by continuously evaluating and filtering content based on its emotional and contextual importance. DIMF prioritizes retention of emotionally significant interactions and contextually relevant information, discarding or further compressing less critical memories.

The DIMF algorithm consists of the following processes:

- **Importance Scoring**: Each new memory entry receives an importance score based on emotional intensity, user feedback, or contextual uniqueness.
- **Periodic Pruning**: Regular evaluations identify and prune entries with importance scores below dynamically determined thresholds, significantly reducing unnecessary memory load.
- **Contextual Preservation**: Important contextual information is maintained through meta-entries or high-level summaries, ensuring key details remain accessible despite aggressive pruning.
- **User Feedback Integration**: The system incorporates user corrections to preserve memories deemed significant by users, further refining memory retention strategies.

By integrating TBC and DIMF, Livia effectively balances detailed memory retention with practical resource constraints, achieving a highly personalized and contextually aware interaction experience (Fig. 3).

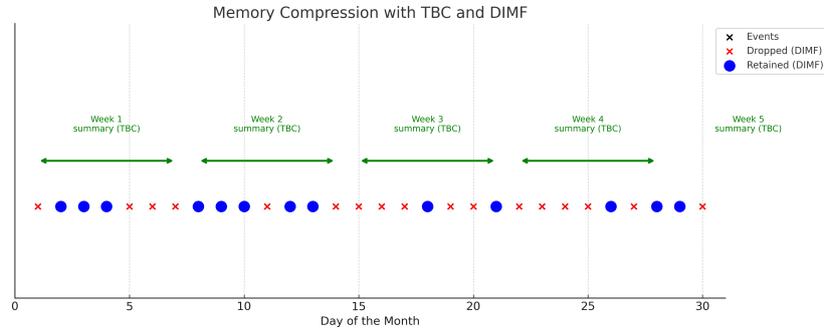

**Fig. 3.** Memory Compression with TBC and DIMF

## 6 Emotion-Aware Engagement System

Livia's Emotion-Aware Engagement System proactively supports users by continuously monitoring their emotional state, leveraging multimodal data, and initiating contextually appropriate interactions. This system significantly extends beyond passive interaction models, providing dynamic and personalized emotional support.

### 6.1 Multimodal Emotion Sensing

To accurately understand user emotions, Livia integrates various sources of data:
- **Text and Dialogue Context**: User utterances are analyzed for sentiment and emotional keywords, while dialogue context offers additional cues about the user's emotional state, such as expressions of loneliness or frustration (Fig. 4).

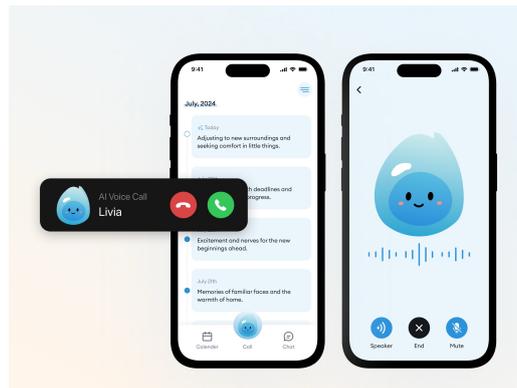

**Fig. 4.** Livia detects and responds to user emotions through intuitive visual and audio feedback.

- **Calendar and Environmental Context**: Livia accesses calendar information (with user consent) to identify emotionally significant upcoming events, such as exams or anniversaries. Temporal patterns (e.g., Monday blues, weekend positivity) are also considered. Future versions may incorporate real-time facial expression analysis during video interactions to further enhance accuracy.

These multimodal data streams enable the Behavior Orchestration Agent to estimate the user's current emotional state and track emotional trends over time, providing a nuanced basis for interaction.

### 6.2 Proactive Dialogue and Contextual Interventions

Using emotion estimation, Livia proactively initiates supportive interactions:

- **Emotional Check-ins**: When sustained negative emotional states, such as stress, are detected, Livia may gently initiate contact through notifications or voice prompts, offering companionship or relaxation techniques (Fig. 4).
- **Contextual Reminders and Suggestions**: Leveraging calendar data, Livia proactively engages the user around significant upcoming events (e.g., job interviews), offering support or preparation activities. Livia also adapts its interaction schedule based on user habits, encouraging healthier behaviors and timely engagement.
- **Adaptive Dialogue Responses**: Even in reactive interactions, Livia utilizes emotional sensing to enrich dialogue quality. For instance, if sensor data contradicts a user's stated emotional state, Livia sensitively prompts further discussion, demonstrating empathy without intrusiveness.

These proactive approaches demonstrate empathetic initiative, improving user engagement by addressing emotional needs promptly and effectively.

### 6.3 Personalization of Emotional Responses

Recognizing the variability in user preferences for emotional interactions, Livia continuously personalizes its engagement strategy based on user feedback and interaction history. User responses to proactive engagements (positive, neutral, or negative) are tracked and utilized to adjust future interaction strategies. For example, if a user frequently responds positively to supportive encouragement, Livia enhances the frequency and enthusiasm of such interactions. Conversely, if a user prefers minimal emotional interaction, Livia will respect this preference by reducing proactive prompts.

This adaptive mechanism ensures personalized, emotionally resonant interactions, fostering a sense of genuine companionship and understanding, significantly enhancing user satisfaction and emotional support effectiveness.

## 7 AR-Driven Embodied Interaction

Livia leverages augmented reality (AR) to create immersive, personalized interactions, significantly enhancing the user's emotional experience and sense of presence. AR tech-nology transforms Livia from a traditional 2D interface into a lifelike companion embedded in the user's physical space, making interactions more authentic and emotionally engaging.

### 7.1 Character Design and Animated Interaction

To achieve a more engaging and realistic presence, we utilized Blender for designing distinctive character avatars and their associated animations, then integrated these into the real-world environment using Unity and ARKit. In AR mode, Livia is rendered as a dynamic, animated avatar that can naturally enter and interact within the user's living space, simulating a tangible friend and enhancing the sense of companionship.

Real-time AI-driven emotion analysis further enriches this experience by dynamically adapting the avatar's expressions and movements in response to the user's current emotional state. This integration between animation and real-time emotional responsiveness significantly increases interaction authenticity and enjoyment.

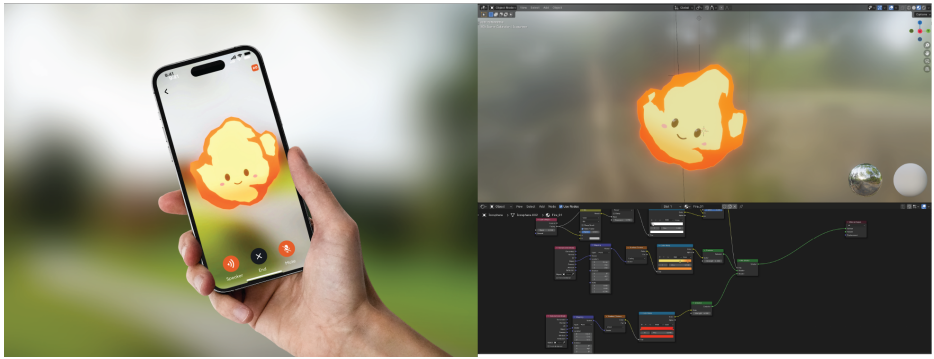

**Fig. 5.** Example of Livia AR Interaction.

### 7.2 Personality Element Matching

To address diverse emotional needs among users, Livia incorporates three distinct personality elements—Fire (energetic and playful), Water (gentle and nurturing), and Earth (steady and reliable). Each personality element features customized visual designs and unique voice characteristics, enabling a deeper and more personalized emotional connection. Users can select and interact with the personality that best aligns with their emotional preferences, significantly enhancing personalization.

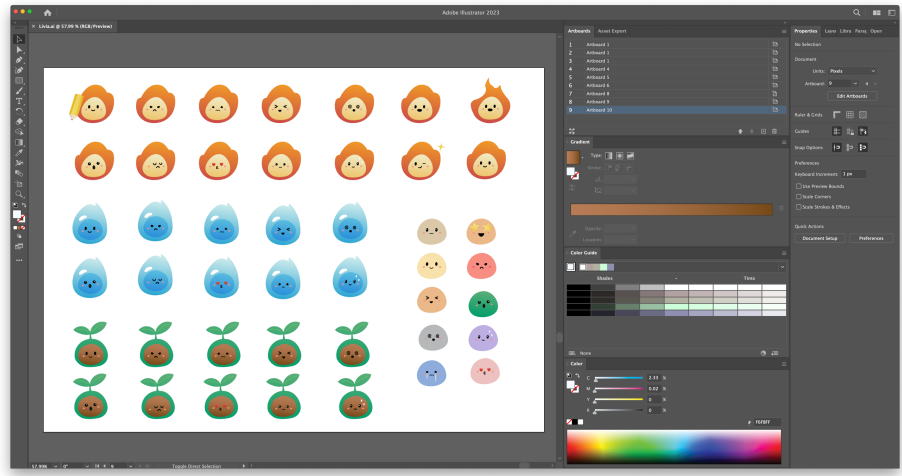

**Fig. 6.** Livia's three personality elements—Fire, Water, and Earth.

### 7.3 Innovative Integration of AR and AI

The integration of AR and AI introduces innovative interaction modes, where Livia becomes an intuitive presence within the user's physical environment. This immersive integration transforms traditional cognitive behavioral therapy (CBT) techniques into engaging, interactive experiences, significantly boosting user participation and the effectiveness of psychological support. AR helps create a believable and comforting environment, while AI provides real-time emotional understanding, making therapeutic interactions enjoyable, accessible, and effective.

## 8 Evaluation

We conducted an extensive evaluation of Livia, focusing on quantitative performance metrics and qualitative user feedback. Our evaluation addresses key questions:
1. How accurately and efficiently does Livia perform core AI tasks, such as emotion recognition, memory compression, response generation, and AR rendering?
2. How do users perceive Livia in terms of emotional support, companionship, and personality depth, and does using Livia correlate with reductions in loneliness?
3. How effectively do memory compression algorithms retain important information while reducing memory size?

To answer these questions, we analyzed user interaction data gathered from backend analytics, conducted user interviews with participants actively using Livia, and carried out in-depth interviews with a selected group of users to obtain richer insights into their experiences.

## 8.1 Quantitative and Qualitative Results

- **Emotion Recognition**: Livia's multimodal emotion detection was evaluated using 200 real chat excerpts from 50 unique users between January and March 2025. Users ranged from 18 to 34 years old (74% female, 26% male), all English speakers from North America. Each chat excerpt was labeled for emotion (happiness, sadness, anxiety, anger, neutral) by two independent human raters (inter-rater agreement: 0.82, Cohen's kappa). Livia's model achieved an overall accuracy of 88%, outperforming a basic text-only version of Livia (75% accuracy). For high-intensity emotions—especially anxiety—adding voice tone improved precision to 92% (vs. 71% with text only).
- **User Engagement**: User engagement was tracked over four weeks with 38 participants. On average, users had 7.9 conversations per day, each lasting about 4.8 minutes. Compared to a standard text-only version of Livia, users interacting with the full multimodal version had 31% higher engagement.
- **Memory Compression**: We tested memory compression algorithms across all user logs (11,504 conversation turns in total). On average, per-user memory storage was reduced from 50KB to 15KB over the evaluation period, without a loss in key recall performance. To measure recall, we defined "important events" as moments users referenced again in later chats. Livia accurately recalled these important events 92% of the time, and less significant details (referenced only once or not brought up again) 65% of the time.
- **AR Rendering and Qualitative Feedback**: All participants used Livia with AR rendering enabled. In structured surveys and follow-up interviews with 24 users, common themes included Livia's perceived empathy, realistic AR visuals, and a strong preference for Livia over text-only chatbots. Several users highlighted the value of Livia's lifelike presence and emotional responsiveness. For example, one user commented, "Livia felt like a real friend who actually remembers what I said yesterday."
- **Limitations**: The main limitations were sample size, demographic homogeneity (all participants were from North America and relatively young), and English-only data. Additionally, emotion labels were assigned by human raters rather than directly by users, which could introduce some subjective bias.

Participants reported strong emotional bonds with Livia and high overall satisfaction. Many described Livia as a comforting presence, highlighting empathy-driven interactions and appealing AR visuals.

# 9 Ethical Considerations and Deployment Challenges

## 9.1 Ethical Considerations

The proliferation of AI companionship raises critical ethical considerations, particularly around emotional reliance. Researchers have expressed concern that prolonged dependence on AI companions could lead to diminished human interaction and emotional deskilling, undermining users' capacities to maintain healthy relationships and manage emotional complexity [10]. Livia directly addresses these concerns by explicitly focusing on a form of companionship analogous to a friend, magical diary, or trusted confidant, deliberately avoiding romantic or erotic dimensions that are often linked to more profound emotional risks. This design philosophy aims to build trustworthy, supportive interactions without inadvertently promoting unhealthy emotional dependencies or substituting critical human relationships [11].

However, even with these safeguards, there remains a risk that users could become overly reliant on Livia, potentially leading to decreased engagement with human social networks. To mitigate this, Livia proactively encourages users to sustain and enhance their existing human relationships, intentionally positioning itself as a supportive resource rather than a replacement for genuine human interaction [12].

Moreover, AI companion applications like Livia inherently offer significant benefits related to inclusivity and diversity. By carefully designing interaction models, developers can overcome biases commonly present in broader AI systems, such as underrepresentation of marginalized groups or insensitivity towards different genders, sexual orientations, ages, or disabilities [13]. Livia prioritizes diverse user needs and contexts, thus fostering an inclusive environment that respects and validates the identities of typically marginalized or overlooked communities. Such intentional inclusivity not only broadens the accessibility of emotional support but also reinforces fundamental human rights by creating spaces that are safe, affirming, and welcoming for all users [14].

Ultimately, the ethical challenge lies in designing AI companions during a time when the long-term effects of human-AI emotional interaction remain uncertain. Although creating AI companions with human-like qualities can amplify both risks and benefits, emphasizing their supportive role in strengthening rather than replacing human relationships can significantly mitigate the risk of social displacement [12]. As with human relationships, when a companion actively encourages nurturing existing social connections, the likelihood of detrimental replacement effects can be substantially reduced [12].

### 9.2 Real-world Deployment Challenges

Deploying an emotionally-aware AR companion like Livia faces substantial practical hurdles in real-world contexts.

Continuous AR rendering presents technical challenges, notably battery drain, overheating, and high computational demands. To address these, efficient rendering algorithms, adaptive frame rates, and optimized resource allocation are essential. Leveraging edge computing and on-device processing can further improve battery life and responsiveness.

Privacy is critically important given the extensive collection of sensitive multimodal user data—including text inputs, voice recordings, facial expressions, and potentially physiological signals. Studies in affective computing emphasize the heightened sensi-tivity and potential risks associated with emotional and psychological data if inadequately protected [15]. To address these concerns, robust data encryption, secure storage protocols, transparent privacy policies, and explicit user consent mechanisms are indispensable. Advanced privacy-preserving technologies can further mitigate data misuse risks, enhance user trust, and support long-term adoption.

The variability in an individual's emotional expression across different moods poses challenges for reliable emotion recognition and adaptive responsiveness. Users exhibit diverse behaviors depending on their emotional context, complicating accurate interpretation by AI systems. Adaptive algorithms capable of continuous personalization and learning from nuanced emotional changes are therefore essential to ensure contextually appropriate and individually tailored interactions.

Addressing these real-world deployment challenges is vital for ensuring the practical feasibility, ethical soundness, and long-term effectiveness of emotionally-aware AR companions like Livia.

## 10    Conclusion & Future Work

Livia significantly advances personalized emotional support by integrating modular AI architectures with immersive AR technologies [9]. Its core contributions include robust multimodal emotion recognition, efficient long-term memory management through progressive compression, and realistic adaptive interactions driven by a responsive personality model. These innovations provide users emotionally intuitive interactions while setting new benchmarks for ethical and inclusive AI companion design.

Future work will explore additional interactive modalities such as gesture-based and tactile interactions to further enhance emotional engagement [16]. Gesture recognition could enable intuitive communication, and tactile feedback might strengthen companionship through simulated physical presence. We also plan to expand Livia to support multi-user and multi-character experiences, promoting richer social interactions and addressing loneliness more effectively. Moreover, integrating Livia into customized hardware solutions like companion dolls or wearable devices will be investigated, presenting opportunities and challenges in form-factor design, ergonomics, and advanced sensory integration.

By pursuing these future directions, Livia is positioned to continually evolve, providing meaningful companionship and emotional support, while also pushing forward the technical and ethical boundaries in AI-driven emotional intelligence and user interaction design.